\documentclass[aps,pra,twocolumn,showpacs]{revtex4}
\usepackage{graphicx}
\usepackage{amsmath}
\usepackage{amssymb}
\newcommand{\ket}[1]{\ensuremath{\left|{#1}\right\rangle}}
\newcommand{\bra}[1]{\ensuremath{\left\langle{#1}\right |}}
\newcommand{\scalprod}[2]{\ensuremath{\left\langle{#1}|{#2}\right\rangle}}
\newcommand{\rmvect}[1]{\boldsymbol{\mathrm{#1}}}

\newcommand{\sinc}{\ensuremath{\mathrm{sinc}}}

\newcommand{\HG}{{\ensuremath{{\mathrm{HG}}}}}
\newcommand{\DHG}{\ensuremath{{\mathrm{DHG}}}}

\newcommand{\lm}{\ell}
\newcommand{\gouy}{\ensuremath{\varepsilon}}

\newcommand{\brm}[1]{\ensuremath{\mathbf{#1}}}

\newcommand{\tr}{\ensuremath{\mathrm{tr}}}
\begin{document}
\title{Conservation and entanglement of Hermite-Gaussian modes in parametric down-conversion}
\author{S. P. Walborn}
\email[]{swalborn@if.ufrj.br}
\altaffiliation[Current address:  ]{Instituto de F\' \i sica, Universidade Federal do Rio de Janeiro, Caixa Postal 68528, Rio de Janeiro, RJ 21945-970, Brazil}
\affiliation{Universidade Federal de Minas Gerais, Caixa Postal 702, Belo Horizonte, MG
30123-970, Brazil}
\author{S. P\'adua}
\affiliation{Universidade Federal de Minas Gerais, Caixa Postal 702, Belo Horizonte, MG
30123-970, Brazil}
\author{C. H. Monken}
\affiliation{Universidade Federal de Minas Gerais, Caixa Postal 702, Belo Horizonte, MG
30123-970, Brazil}
\date{\today}
\begin{abstract}
We show that the transfer of the angular spectrum of the pump beam to the two-photon state in spontaneous parametric down-conversion enables the generation of entangled Hermite-Gaussian modes.  We derive an analytical expression for the  two-photon state in terms of these modes and show that there are restrictions on both the parity and order of the down-converted Hermite-Gaussian fields.  Using these results, we show that the two-photon state is indeed entangled in Hermite-Gaussian modes.  We propose experimental methods of creating maximally-entangled Bell states and non-maximally entangled pure states of first order Hermite-Gaussian modes.
\end{abstract}
\maketitle
\section{Introduction}
Recently, a great deal of attention has been paid to the higher-order Gaussian modes of the electromagnetic field.  In the paraxial approximation, two interesting cases are the Laguerre-Gaussian (LG) and Hermite-Gaussian (HG) modes.  These modes are solutions of the paraxial Helmholtz equation \cite{saleh91} and are eigenstates of the free-space propagator.  It has been shown that the LG modes carry orbital angular momentum in the form of an azimuthal phase $e^{i \lm \phi}$ in the transverse
plane \cite{allen92,vanenk92}.  Allen \textit{et al.} \cite{allen92} have shown that field modes with this type of phase dependence carry an orbital angular momentum of $\lm \hbar$ per photon, where $\lm$ is the azimuthal beam index.
\par
These higher-order modes are of great interest in quantum information schemes, since they can be used to represent discrete $\mathcal{D}$-state \emph{qudits}.  For example, the orbital angular momentum of single photons in LG modes provides a possible qudit encoding scheme.  The quantum number $\lm$ can be coherently raised or lowered using holographic masks \cite{heckenberg92}.   One can measure the orbital angular momentum (LG modes) of single photons up to a desired accuracy using interferometric techniques \cite{leach02,wei03_OC117}.    In, addition,   Mair \textit{et al.} \cite{mair01} have shown experimentally that with spontaneous parametric down-conversion (SPDC) it is possible to create photon pairs entangled in orbital angular momentum and other theoretical \cite{arnaut01,franke-arnold02,barbosa02,torres03_PRA052313,walborn04a,ren04,law04} and experimental works \cite{walborn04a} have followed, including the generation of entangled 3-state qutrits \cite{vaziri02,langford03}.
\par
The HG modes may also be of use in quantum information schemes.  In particular, the first-order HG and LG modes can be described and manipulated in a way that is analogous to linear and circular polarization of the electromagnetic field \cite{oneil00}.  Devices that act on the first-order transverse mode in a manner equivalent to polarizing beam splitters, half-wave plates and quarter-wave plates can be constructed using asymmetric interferometers \cite{xue01_OL,sasada03_PRA012323}, Dove prisms \cite{oneil00} and mode converters \cite{beijersbergen93}.  The first-order modes can thus be used to define qubits and the above devices implement single qubit rotations.  Recently, Langford \textit{et al.} have produced photons  entangled in first-order HG mode and performed quantum state tomography using holographic masks and single mode fibers \cite{langford03}.  Since the HG modes form an infinite-dimensional orthonormal basis, they too might be used to encode higher-dimensional qudits.
\par
Here we provide a theoretical description of the generation of entangled HG modes for arbitrary HG pump beams.  We show that there are restrictions on the parity and order of the down-converted HG fields.  We introduce our notation in section \ref{sec:hg} and briefly review the two-photon quantum state generated by SPDC in section \ref{sec:state}.  Our main results concerning the generation of correlated HG modes using SPDC are derived in section \ref{sec:gen}, including a general expression for the probability amplitude to generate combinations of different HG modes.  In section \ref{sec:ent}, we provide a proof that the down-converted HG modes are indeed entangled and we discuss the experimental generation of Bell-states and non-maximally entangled pure states.
\subsection{HG modes}
\label{sec:hg}
 For convenience, we will adopt the notation used in Ref. \cite{beijersbergen93}. The Hermite-Gaussian modes are given by the complex field amplitude
\begin{align}
\HG_{nm}(x,y,z)=&
C_{nm}\frac{1}{w(z)}
H_{n}\left(\frac{\sqrt{2}x}{w(z)}\right)H_{m}\left(\frac{\sqrt{2}y}{w(z)}\right) \times \nonumber \\ & \exp\left(-\frac{x^2+y^2}{w(z)^2}
\right) \times \nonumber \\ & \exp\left[-i\frac{k(x^2+y^2)}{2R}
-i(n+m+1)\gouy(z)\right],
\label{eq:hg}
\end{align}
where the coefficients $C_{nm}$ are given by
\begin{equation}
C_{nm} = \sqrt{\frac{2}{2 ^{(n+m)} \pi n! m!}},
\label{eq:Cnm}
\end{equation}
and $H_{n}(x)$ is the $n^{\mathrm{th}}$-order Hermite polynomial.  The radius of curvature $R(z)$, beam waist $w(z)$ and Gouy phase $\gouy(z)$
are given by
\begin{equation}
w(z)=w_{0}\sqrt{1+\frac{z^{2}}{z_{R}^{2}}},
\label{eq:gauss2}
\end{equation}
\begin{equation}
R(z)=z\left({1+\frac{z^{2}}{z_{R}^{2}}}\right),
\label{eq:gauss3}
\end{equation}
and
\begin{equation}
\gouy(z)=\arctan\frac{z}{z_{R}},
\label{eq:gauss4}
\end{equation}
respectively. The parameter $z_{R}$ is the
Rayleigh range.  The order $\mathcal{N}$ of the beam is the sum of the indices $\mathcal{N} = m+n$.  Note that the usual Gaussian beam is the zeroth-order $\HG_{00}$ beam.
\par
In section \ref{sec:gen}, we will make use of the diagonal Hermite-Gaussian modes (DHG) defined by \cite{beijersbergen93,oneil00}
\begin{equation}
\DHG_{nm}\left(\tilde{x},\tilde{y},z \right) =\sum\limits_{k=0}^{n+m}b(n,m,k)\HG_{N-k,k}(x,y,z),
\label{eq:DHGexpan}
\end{equation}
with $\tilde{x} = (x+y)/{\sqrt{2}}$, $\tilde{y} = (x-y)/{\sqrt{2}}$ and the coefficient $b(n,m,k)$ defined as
\begin{equation}
 b(n,m,k) = \sqrt{\frac{(n+m-k)!k!}{2^{(n+m)}n!m!}}\frac{1}{k!}\frac{d^{k}}{dt^{k}}\left[(1-t)^{n}(1+t)^{m} \right]\big|_{t=0}.
 \label{eq:b}
\end{equation}
 \section{State generated by SPDC}
 \label{sec:state}
 Here we review the two-photon quantum state generated by SPDC.  We consider that a photon $p$ from a sufficiently weak cw pump beam is incident on a nonlinear crystal,  producing down-converted signal and idler photons $s$ and $i$, respectively.  We will work in the monochromatic approximation, which is justified experimentally by the use of narrow bandwidth interference filters in the detection system.  It is assumed that the filters are centered at the degenerate wavelength $\lambda_{c}=2 \lambda_{p}$, where $\lambda_{p}$ is the pump beam wavelength. We will also work in the paraxial approximation, which will be discussed below.  For a sufficiently weak cw laser, the quantum state generated by
SPDC can be written as \cite{hong85,monken98a}
\begin{equation}
\ket{\psi}_{12} = C_{1}\ket{\mathrm{vac}} + C_{2}\ket{\psi},
\end{equation}
where
\begin{equation}
\ket{\psi}=\sum_{\sigma_{s},\sigma_{i}}C_{\sigma_{s},\sigma_{i}}\int\hspace{-2mm}\int\limits_{D}\hspace{-1mm} d\brm{q}_{s}
d\brm{q}_{i}\ \Phi(\brm{q}_{s},\brm{q}_{i})\ket{\brm{q}_{s},\sigma_{s}}_{s}
\ket{\brm{q}_{i},\sigma_{i}}_{i}.
\label{eq:state}
\end{equation}
The coefficients $C_1$ and $C_2$ are such that $|C_{2}| \ll \, |C_{1}|$. $C_2$ depends on the nonlinearity coefficient and length of the nonlinear crystal, the magnitude of the pump beam, as well as other experimental parameters. The ket $\ket{\brm{q}_{j},\sigma_{j}}$ represents  a single-photon state in a plane wave mode. The vector $\brm{q}_{j}$ is the transverse component of the wave vector
$\brm{k}_{j}$ and $\sigma_{j}$ is the polarization of the mode $j = s$ or $i$, where the sum is over two orthogonal polarization directions $\sigma_{j}$ and  $\bar{\sigma}_{j}$. The polarization state of the down-converted photon pair is defined by the coefficients $C_{\sigma_{s},\sigma_{i}}$. The normalized function $\Phi(\brm{q}_{s},\brm{q}_{i})$ is given by \cite{monken98a}
\begin{equation}
\Phi(\brm{q}_{s},\brm{q}_{i}) =\frac{1}{\pi}\sqrt{\frac{2L}{K}}\
v(\brm{q}_{s}+\brm{q}_{i})\
\sinc\left(\frac{L|\brm{q}_{s}-\brm{q}_{i}|^{2}}{4K} \right),
\label{eq:state2}
\end{equation}
where $v(\brm{q})$ is the normalized angular spectrum of the pump beam, $L$ is the length of the nonlinear crystal in the propagation ($z$) direction, $\sinc(x) \equiv (\sin x)/x$, and $K$ is the magnitude of the pump field wave vector. The integration domain $D$ is defined as the region in which the paraxial approximation is valid.  In most experimental conditions, however, $D$ is much larger than the region in which $\Phi(\brm{q}_{s},\brm{q}_{i})$ is appreciable.
\par
  We assume that $\Phi(\brm{q}_{s},\brm{q}_{i})$ does not depend on the polarizations of the down-converted photons. This assumption may not be true, especially if the crystal is cut for type-II SPDC.  However, the polarization dependence can be reduced by placing birefringent crystal compensators in the down-converted beams \cite{kwiat95}.
\par
We note here that recent experimental work \cite{monken98a,atature02_PRA23822,walborn04a} has shown that the quantum state \eqref{eq:state} is an accurate description of the two-photon component of the quantum state generated by SPDC using a cw laser.
\section{Generation of entangled HG modes with SPDC}
\label{sec:gen}
\begin{figure}
 \begin{picture}(220,100)(0,0)
\put(10,10){\includegraphics[width=7cm]{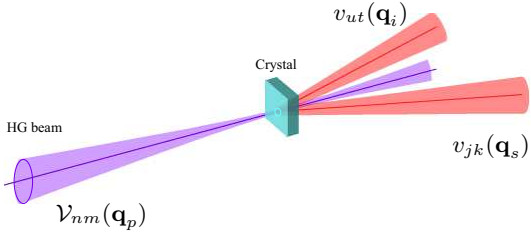}}%
  \put(30,4){$\mathcal{V}_{nm}(\rmvect{q}_{p})$}
   \put(135,80){$v_{ut}(\rmvect{q}_{i})$}
    \put(180,30){$v_{jk}(\rmvect{q}_{s})$}
 \end{picture}
 \caption{\label{fig:setup} The angular spectrum of the pump beam $\mathcal{V}_{nm}(\rmvect{q}_{p})$, characterized by the wavelength $\lambda_{p}$ and beam width the $w_{0p}$, creates down-converted fields with angular spectra $v_{jk}(\rmvect{q}_{s})$ and $v_{ut}(\rmvect{q}_{i})$, characterized by the wavelength $\lambda_{c} = 2\lambda_{p}$ and the beam width $w_{0c}=\sqrt{2}w_{0p}$.}

 \end{figure}

In the following we will denote $v_{nm}(\rmvect{q})$ as the normalized angular spectrum of the HG mode, which can be calculated by taking the two-dimensional Fourier transform of (\ref{eq:hg}).  Explicitly,
   \begin{align}
v_{nm}(q_{x},q_{y}) = & w D_{nm}H_{n}\left(\frac{w q_{x}}{\sqrt{2}}\right)H_{m}\left(\frac{w q_{y}}{\sqrt{2}}\right) \times \nonumber \\
& \exp\left(-\frac{w^2(q_{x}^2+q_{y}^2)}{4}
\right) \times \nonumber \\ & \exp\left[-i
(n+m+1)\gouy(z)\right],
\label{eq:hgspdc6}
\end{align}
where $w \equiv w(z)$ and
\begin{equation}
D_{nm} =  \frac{-i^{(n+m)}}{2}C_{nm}
\label{eq:Dnm}
\end{equation}
such that $v_{nm}(q_{x},q_{y})$ is properly normalized.
The general problem we are considering is illustrated in FIG. \ref{fig:setup}.  We now consider that the non-linear crystal is pumped with a Hermite-Gaussian beam $\HG_{nm}$, generating a two-photon state \ket{\psi_{nm}}.  To account for the different wavelengths of the pump and down-converted fields, we will write the angular spectrum of the HG pump beam as $\mathcal{V}_{nm}$, which is equivalent to expression (\ref{eq:hgspdc6}) but characterized by the wavelength $\lambda_{p}$ and the beam radius $w_{0p}$.  The angular spectrum of the down-converted field $v_{\alpha\beta}$ is characterized by the wavelength $\lambda_{c}$ and the beam radius $w_{0c}$.  It will be shown in the appendix that $w_{0c}=\sqrt{2}w_{0p}$.

 \par
 Since the HG beams form a complete basis, we can expand the two-photon state as
\begin{equation}
\ket{\psi_{nm}} = \sum_{j,k,u,t=0}^{\infty}\{ {}_{i}\hspace{-0.8mm}\bra{v_{ut}}{}_{s}\hspace{-0.8mm}\scalprod{v_{jk}}{\psi_{nm}} \}
\ket{v_{jk}}_{s}\ket{v_{ut}}_{i},
\end{equation}
where we have introduced the shorthand notation
\begin{equation}
\ket{v_{\alpha\beta}} = \int d\rmvect{q} v_{\alpha\beta}(\rmvect{q}) \ket{\rmvect{q}}.
\label{eq:vket}
\end{equation}
We note here that $j$ ($k$) and $u$ ($t$) are the $x$ ($y$) indices of the signal and idler fields, respectively.  To facilitate the calculations, we will assume that $z_{s}=z_{i}=0$ at the crystal face.  Defining
\begin{equation}
C^{\,nm}_{jkut} = \bra{v_{ut}}\scalprod{v_{jk}}{\psi_{nm}},
\label{eq:Cjkst}
\end{equation}
we have
\begin{equation}
\ket{\psi_{nm}} = \sum_{j,k,u,t=0}^{\infty}C^{\,nm}_{jkut}\ket{v_{jk}}\ket{v_{ut}}.
\label{eq:stateHG}
\end{equation}
The task at hand is to calculate the coefficients $C^{\,nm}_{jkut}$.
\par
For simplicity, we assume that the down-converted fields are not entangled in polarization.  Then, we can ignore the polarization dependence of the two-photon state \eqref{eq:state}.  Depending on the type of phase matching,  the pump and down-converted fields may suffer a small astigmatism when propagating through the birefringent non-linear crystal \cite{costamoura04}.  This astigmatism depends on the order of the modes as well as the length $L$ of the non-linear crystal, being negligible for thin crystals and/or low-order modes.  Here we will assume that the crystal is cut for type-I phase matching such that the pump beam is polarized in the extraordinary direction and suffers an astigmatism, while the ordinarily polarized down-converted fields do not suffer any deformation.  We will also assume that the pump beam is of low order $\mathcal{N}=n+m \leq 2$ and consider that the crystal length is on the order of a few millimeters.  Under these conditions, we can ignore the birefringence and astigmatism effects \cite{costamoura04}.  Then, using Eqs. \eqref{eq:state}, \eqref{eq:state2} and \eqref{eq:vket} in Eq. \eqref{eq:Cjkst} gives
\begin{align}
C^{\,nm}_{jkut} = & \frac{1}{\pi}\sqrt{\frac{2 L}{K}} \iint    d\rmvect{q}_{s}\,d\rmvect{q}_{i} v^{*}_{jk}(\rmvect{q}_{s})v^{*}_{ut}(\rmvect{q}_{i}) \times \nonumber \\ & \mathcal{V}_{nm}(\rmvect{q}_{s}+\rmvect{q}_{i})\,
\sinc \left(\frac{L}{4K}|\rmvect{q}_{s}-\rmvect{q}_{i}|^2 \right ).
\label{eq:hgspdc7}
\end{align}
In the appendix, we show that
\begin{align}
C^{\,nm}_{jkut} = & \sqrt{\frac{\alpha!\beta! }{A \pi}}\left( \frac{1}{2} \right)^{\frac{\alpha+\beta}{2}} \frac{\arctan A}{(\alpha/2)!(\beta/2)!} b(j,u,\alpha)b(k,t,\beta) \times \nonumber \\
& \sum\limits_{r=0}^{(\alpha+\beta)/2}
\left (\begin{array}{c}
 \frac{\alpha+\beta}{2} \\  r
\end{array} \right)
\left( \frac{-2}{\sqrt{1+A^2}}\right)^{r} \sinc ( r \arctan A)
 \label{eq:Cexact}
\end{align}
if $j+u \geq n$ and $k+t \geq m$, else $C^{\,nm}_{jkut} = 0$.  Here $b(j,u,\alpha)$ is given by Eq. \eqref{eq:b} and we have defined $A=L/Kw_{p}^2$, $\alpha=N-n$ and $\beta=M-m$.
 \par
In the appendix, it is also shown that for thin non-linear
crystals ($L \sim 1$ mm), ${C}^{\,nm}_{jkut}$ simplifies to
\begin{align}
{C}^{\,nm}_{jkut}\rightarrow {\tilde{C}}^{\,nm}_{jkut} =
  & \sqrt{\frac{2 A}{\pi}}b(j,u,N-n)b(k,t,M-m) \times \nonumber \\
  & \HG_{N-n,M-m}(0,0,0),
    \label{eq:Camp}
 \end{align}
 if $j+u \geq n$ and $k+t \geq m$, otherwise $C^{\,nm}_{jkut} = 0$.    Here $\HG_{\gamma\delta}$ is the Hermite-Gaussian mode (\ref{eq:hg}) evaluated at $x=y=z=0$.  When $\gamma$ is odd, the Hermite polynomial $H_{\gamma}(0)=0$, which gives another conservation condition:  $N-n$ and $M-m$ must be even.  In other words, the sum of the $x$ ($y$) indices of the down converted fields $N=j+u$ ($M=k+t$) must have the same parity as the $x$ ($y$) index of the pump field $n$ ($m$).  In summary, the conservation conditions are
\begin{subequations}
\label{eq:HGcons}
 \begin{align}
 j+u & \geq n  \hspace{2mm} \mathrm{and} \hspace{2mm} \mathrm{parity}\,(j+u) = \mathrm{parity}\,n,  \\
 k+t & \geq m  \hspace{2mm} \mathrm{and} \hspace{2 mm} \mathrm{parity}\,(k+t) = \mathrm{parity}\,m.
 \end{align}
  \end{subequations}
  We note here that the conservation conditions restrict, for example,  the sum $j+u$ and not $j$ or $u$ individually.
  \par
 The parity of the product of the signal and idler HG
modes (or the sum of the HG mode indices) must maintain the parity
of the pump beam angular spectrum, which has been transferred to
the two-photon quantum state. From a mathematical point of view,
these results are intuitive. For example, consider an even
function $f$ and an expansion of the sort
\begin{equation}
f(x+y) = \sum\limits_{i} A_{i} g_{i}(x) h_{i}(y).
\end{equation}
The even parity of $f$ requires that $f(x+y) = f(-x-y)$ or
\begin{equation}
\sum\limits_{i} A_{i} g_{i}(x) h_{i}(y) = \sum\limits_{i} A_{i}
g_{i}(-x) h_{i}(-y).
\end{equation}
Since $f$ is an even function, all products $g_{i}(x) h_{i}(y)$ in
the expansion must have the same parity, which in this example
implies that either $g$ and $h$ are both even function or $g$ and
$h$ are both odd functions.  From the point of view of physics,
the underlying physical process governing the generation of HG
modes is the transfer of the angular spectrum of the pump beam to
the two-photon state (\ref{eq:state}), upon which the derivation
of the coefficients $C^{\,nm}_{jkut}$ and parity and order
restrictions above are based. We note here that it is also the
angular spectrum transfer which is responsible for the generation
of entangled orbital angular momentum states
\cite{franke-arnold02,walborn04a}.
 \par
We have calculated the exact and approximate probability amplitudes for the generation of any arbitrary combination of HG modes with SPDC.  These results show that the indices of the HG modes must obey the conditions (\ref{eq:HGcons}).  Eqs. (\ref{eq:Cexact}), (\ref{eq:Camp}) and (\ref{eq:HGcons}) are the principal results of this paper.  Let us now analyze these results for some particular HG pump beams with typical experimental parameters.
 \begin{figure}
 \includegraphics[width=8cm]{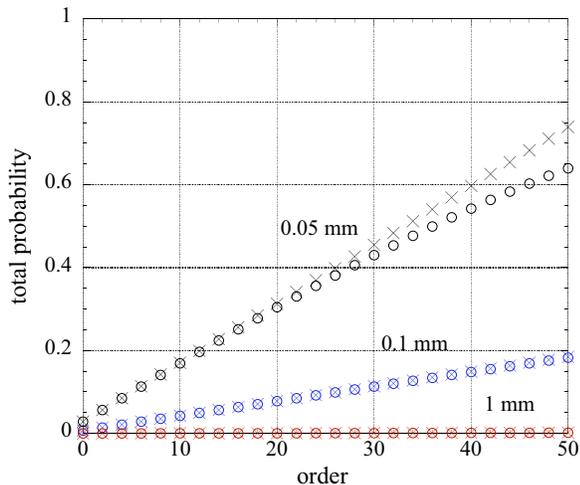}%
 \caption{\label{fig:tot_prob} Total probability of HG mode generation as a function of the order $j+k+u+t$ for a Gaussian pump beams with width $w_{0p}=1$\,mm, $0.1$\,mm and $0.05$\,mm.}
 \end{figure}
  \par
  Fig. \ref{fig:tot_prob} shows the total probabilities obtained by summing all the exact $|C^{00}_{jkut}|^{2}$ (circles) or approximate $|\tilde{C}^{00}_{jkut}|^{2}$ (crosses) probabilities up to a given order $\mathcal{O}=j+k+u+t$.  The pump beam is a Gaussian ($n=m=0$) with $\lambda_{p}=351$\,nm and the crystal length $L$ is $1$\,mm.  Results are shown for pump beam width $w_{0p}=1$\,mm, $0.1$\,mm and $0.05$\,mm.  The total probability $\sum |C^{00}_{jkut}|^{2}$ approaches unity faster for narrower width pump beams.  This indicates that  experimentally one can increase the generation efficiency of lower order modes by focusing the pump beam at the plane of the nonlinear crystal.
 For smaller $w_{0p}$, the approximate solution (\ref{eq:Camp}) is valid only for lower orders.  Calculations of the total probability for an extremely focused pump beam (not shown) shows that the total probability for the exact solution converges to $1$, which indicates that the two-photon state $\ket{\psi_{nm}}$ is properly normalized.
 \par
  The parameter of interest is $A=L/Kw_{0p}^2$, which shows that the generation efficiency of lower-order modes can also be increased by using a longer crystal.  However, we again emphasize that the pump and down-converted fields may suffer greater astigmatic effects in longer crystals.  It is interesting to note that $A$ can also be written as $A=L/(2z_{R})$, where $z_{R}$ is the Rayleigh range of the pump beam.  This shows that the critical parameter is the crystal length $L$ compared to the Rayleigh range of the pump beam.  $\arctan A$ in (\ref{eq:Cexact}) can be viewed as a phase retardation, similar to the Gouy phase (\ref{eq:gauss4}).
  \par
\begin{figure}
  \begin{picture}(230,440)(0,0)
    \put(0,100){$C^{\,11}_{jkut}$}
    \put(0,326){$C^{\,00}_{jkut}$}
\put(20,0){\includegraphics[width=7.5cm]{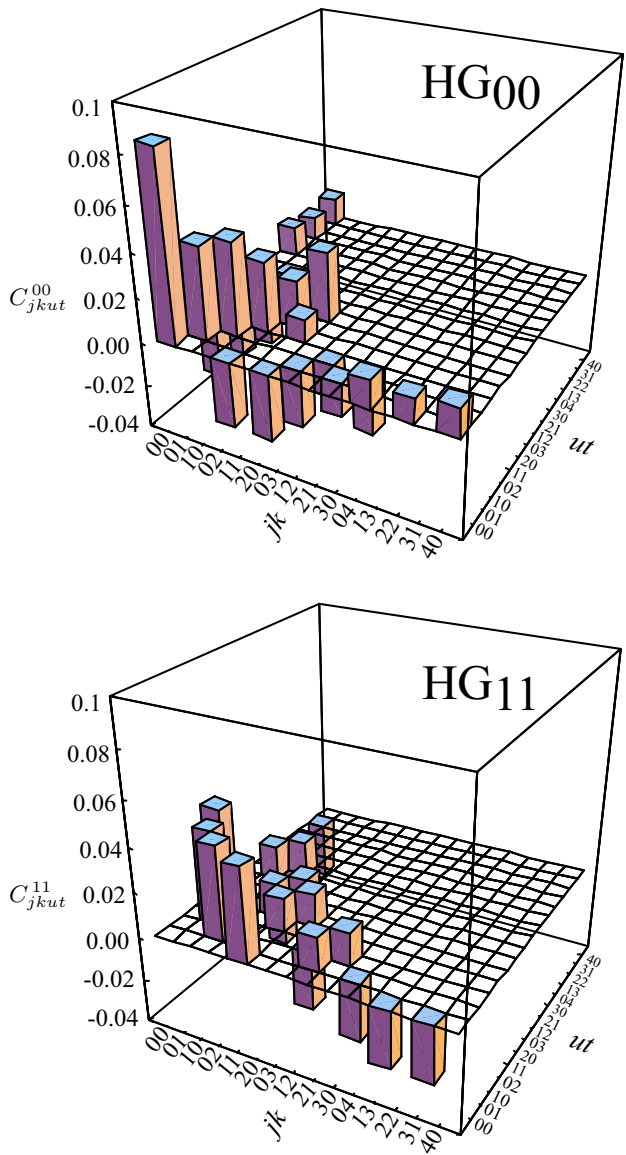}}%
\end{picture}
 \caption{\label{fig:hg00hg11} Coefficients $C^{\,nm}_{jkut}$ up to order $\mathcal{O}=j+k+u+t=4$ for for HG$_{00}$ and HG$_{11}$ pump beams with width $w_{0p}=0.1$\,mm and crystal length $L=1$\,mm.  To improve visualization, only non-zero coefficients have been included.}
 \end{figure}

 Fig. \ref{fig:hg00hg11} contains the amplitudes $C^{00}_{jkut}$ and  $C^{11}_{jkut}$ up to fourth order ($\mathcal{O}=4$) for HG$_{00}$ and HG$_{11}$ pump beams with
 crystal length $L=1$\,mm and pump beam width $w_{p}=0.1$\,mm.  For visual clarity, only non-zero terms have been included.
\section{Entanglement}
\label{sec:ent}
Previous experiments have shown that the pure state given by Eq. (\ref{eq:state}) is an accurate description of the two-photon state generated by SPDC \cite{monken98a,atature02_PRA23822,walborn04a}.
Here we have show that the two-photon state \eqref{eq:state} can also be written as a combination of correlated HG modes in the form \eqref{eq:stateHG} with the normalized coefficients $C^{\,nm}_{jkut}$ satisfying the restrictions on parity and order given in \eqref{eq:HGcons}.  We will now use these restrictions to show that the two-photon state is entangled in HG modes.
\par
 Let us denote the reduced density operator of - say - the signal photon by $\rho_{s}$.  It is well known that $\rho_{s}$  has the following properties \cite{chuang00}:  (\textit{i}) $\rho_{s}$ is a positive operator, (\textit{ii}) $\tr\rho_{s}=1$ and (\textit{iii}) $\tr\rho_{s}^2 \leq 1$ .  If $\rho_{s}$ represents a pure state, then $\tr\rho_{s}=1$, while $\tr\rho_{s}^2<1$ indicates that $\rho_{s}$ represents a mixed state \cite{chuang00}.  For the pure two-photon state $\ket{\psi_{nm}}$,  $\tr\rho_{s}^2 <1$ implies that the overall state is entangled \cite{ballentine98}.  We will show that $\ket{\psi_{nm}}$ is entangled by proving that $\tr\rho^{2}<1$.
\par
It is straightforward to calculate the reduced density operator from the two-photon state (\ref{eq:stateHG}):
 \begin{equation}
 \rho_{s} = \sum\limits_{jkdf=0}^{\infty} F_{jkdf} \ket{v_{jk}}_{s\,s}\hspace{-0.5mm}\bra{v_{df}},
 \label{eq:rho}
 \end{equation}
 where
 \begin{equation}
 F_{jkdf} = \sum\limits_{\gamma\delta=0}^{\infty}C_{jk\gamma\delta}^{\,nm}C_{df\gamma\delta}^{\,nm}.
 \label{eq:Fjkdf}
 \end{equation}
Here we have recognized that the coefficients $C_{jkut}^{nm}$ given by Eq. (\ref{eq:Cexact}) are real. For the argument below, we note that: $F_{jkdf}=F_{dfjk}$, $F_{jkjk}\geq0$ and $\tr\rho_{s}=\sum_{jk=0}^{\infty}F_{jkjk}=1$.  The reduced density operator $\rho_{s}$ satisfies
 \begin{equation}
 \tr\rho_{s}^{2} = \sum\limits_{jkdf=0}^{\infty} (F_{jkdf})^2\leq 1.
 \label{eq:trrho2}
 \end{equation}
 Since $\rho_{s}$ has unity trace, we can write
 \begin{equation}
 \sum_{jk=0}^{\infty}F_{jkjk} \sum_{df=0}^{\infty}F_{dfdf}=1,
 \label{eq:Fid}
 \end{equation}
 so that from Eqs. (\ref{eq:trrho2}) and (\ref{eq:Fid}) we obtain
  \begin{equation}
  \sum\limits_{jkdf=0}^{\infty}\left [ (F_{jkdf})^2 -F_{jkjk}F_{dfdf}\right ] \leq 0.
  \label{eq:Fcond}
  \end{equation}
 $\rho_{s}$ is a positive operator, so its elements satisfy the generalized Cauchy-Schwartz-Buniakowski inequality \cite{kantorovich82}\footnote{Noting that $F_{jkdf}=\bra{v_{jk}}\rho_{s}\ket{v_{df}}$, replacing $\rho_{s}$ with the identity operator gives the usual Cauchy-Schwartz inequality.}:
  \begin{equation}
  (F_{jkdf})^2 \leq F_{jkjk}F_{dfdf}.
  \label{eq:csb}
  \end{equation}   Eq. (\ref{eq:csb}) implies that if $(F_{jkdf})^2 < F_{jkjk}F_{dfdf}$ for any particular values of $j,k,d$ and $f$, then the equality in (\ref{eq:Fcond}) must be false, which indicates that $\tr\rho_{s}^2<1$ and $\ket{\psi_{nm}}$ is entangled.  From the parity conservation conditions (\ref{eq:HGcons}), we see that for any $\gamma$ in the summation in Eq. (\ref{eq:Fjkdf}), $j+\gamma$ and $d+\gamma$ must have the same parity as $n$, otherwise $C_{jk\gamma\delta}C_{df\gamma\delta}=0$.  A similar relation exists for $k+\delta$, $f+\delta$ and $m$.  This implies that $F_{jkdf}=0$ unless $j$ and $d$ \emph{and} $k$ and $f$ have the same parity.  The conditions (\ref{eq:HGcons}) restrict the parity of the sum $j+\gamma$ but not $j$ independently, so $j$ can be either even or odd, as is seen in Fig. \ref{fig:hg00hg11} for the particular cases of $\HG_{00}$ and $\HG_{11}$ pump beams.  Then there exist $F_{jkjk} \neq 0$ and $F_{dfdf} \neq 0$ such that $j$ and $d$ \emph{or} $k$ and $f$ do not have the same parity, which, using the fact that in this case $F_{jkjk} > 0$, implies that $(F_{jkdf})^2=0 < F_{jkjk}F_{dfdf}$.  Then equality in (\ref{eq:Fcond}) is false, which shows that $\ket{\psi_{nm}}$ is entangled.
  \par
  For an infinite dimensional space there will be an infinite number of terms which satisfy the above conditions.  The above proof can also be used to show that the state \ket{\psi_{nm}} is entangled in an arbitrarily large but finite dimensional space, as long as the coefficients (\ref{eq:Cexact}) are properly normalized.  Experimentally, one can post-select the desired HG components of the two-photon state, as will be briefly discussed in the next section.  As long as the reduced density matrix contains one term $F_{jkdf}$ such that $j$ and $d$ \emph{or} $k$ and $f$ have different parity, the equality in (\ref{eq:Fcond}) is  false.
  \subsection{Generating Bell states}
 Through post-selection, it is possible to obtain finite-dimensional entangled states of higher-order Gaussian modes.  Experimentally, post-selection can be achieved by coupling the down-converted fields into optical fibers \cite{mair01,vaziri02,vaziri03a,langford03}, which filter out unwanted modes.  Similarly, entanglement concentration of LG modes was achieved by properly coupling these modes into optical fibers \cite{vaziri03a}.
 \par
  Referring to Fig. \ref{fig:hg00hg11} for the $\HG_{00}$ pump beam, if one considers only first-order down-converted fields ($j+k=1$, $u+t=1$), the resulting quantum state is maximally entangled, resembling the $\phi^{+}$ Bell state, as was observed in \cite{langford03}.  It is then fairly straightforward to experimentally generate all four Bell states using first-order HG modes.  Using a Dove prism (aligned at $45^{\circ}$) to rotate $\HG_{01} \longleftrightarrow \HG_{10}$  of either  the signal or idler field, one can generate the $\psi^{+}$ Bell state.  Placing one additional mirror reflection (or a Dove prism aligned at $90^{\circ}$) in either the signal or idler path, such that $\HG_{01} \longrightarrow -\HG_{01}$ and $\HG_{10} \longrightarrow \HG_{10}$, one can then generate the maximally entangled $\phi^{-}$ and $\psi^{-}$ states.
  \par
  Another method of generating Bell-states of first-order HG modes is with the second-order pump beam $\HG_{11}$.  Isolating only first-order modes, the output state resembles the maximally-entangled $\psi^{+}$ state, as seen in Fig. \ref{fig:hg00hg11}.  This method may be advantageous since the high-probability zero-order $\HG_{00}$-$\HG_{00}$ term is not present.

  \subsection{Generating non-maximally entangled states}
    \begin{table}
   \begin{center}
 \begin{tabular}{|c|c|c|c|c|c|c|}
   \hline
$\mathcal{O}$ & $jk$ &  $ut$   & $C_{jkut}^{02}$  &
$|C_{jkut}^{02}|^{2}$ & $\tilde{C}_{jkut}^{02}$  &
$|\tilde{C}_{jkut}^{02}|^{2}$\\  \hline \hline
2 & 00 & 02 & 0.042169 & 0.001778 & 0.042170 & 0.001778\\
2 & 01 & 01 & 0.059636 & 0.003556 & 0.059637 & 0.003557\\
2 & 02 & 00 & 0.042169 & 0.001778 & 0.042170 & 0.001778\\
\hline\hline
$\mathcal{O}$ & $jk$ &  $ut$   & $C_{jkut}^{20}$  & $|C_{jkut}^{20}|^{2}$ & $\tilde{C}_{jkut}^{20}$  & $|\tilde{C}_{jkut}^{20}|^{2}$\\
\hline \hline
2 & 00 & 20 & 0.042169 & 0.001778 & 0.042170 & 0.001778\\
2 & 10 & 10 & 0.059636 & 0.003556 & 0.059637 & 0.003557\\
2 & 20 & 00 & 0.042169 & 0.001778 & 0.042170 & 0.001778\\
\hline
\end{tabular}
 \caption{\label{tab:hg02} Amplitudes and  probabilities for Hermite-Gaussian pump beams $\HG_{02}$ (top) and $\HG_{20}$ (bottom) up to second order for crystal length $L=1$\,mm and beam width $w_{p}=0.1$\,mm.  The order is defined as $\mathcal{O}=j+k+u+t$.}
\end{center}
 \end{table}
 \begin{figure}
 \includegraphics[width=7cm]{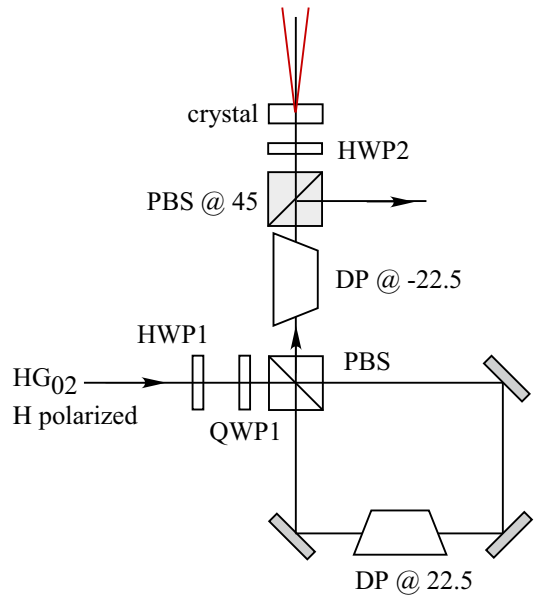}%
 \caption{\label{fig:nonmax} Possible experimental setup to generate non-maximally entangled states (see text).  The input pump beam is a vertically polarized $\HG_{02}$ beam.  A half- (HWP1) and quarter-wave plate (QWP1) are used to adjust the pump polarization.  The Dove prisms (DP) rotate the transverse spatial profile of the pump beam.}
 \end{figure}
  Table \ref{tab:hg02} shows results up to second-order for $\HG_{02}$ and $\HG_{20}$ beams.  Looking at first order terms only ($\HG_{01}$ and $\HG_{10}$), the $\HG_{02}$ pump beam creates an $\HG_{01}-\HG_{01}$ term, while a $\HG_{20}$ pump generates the $\HG_{10}-\HG_{10}$ term.  Creating a pump beam that is an arbitrary coherent superposition of these two beams, we can generate non-maximally entangled pure states.   Fig. \ref{fig:nonmax} shows a possible experimental setup.  The input pump beam is a horizontally polarized $\HG_{02}$ beam.  A half- (HWP1) and quarter-wave plate (QWP1) are used to adjust the pump polarization.  Rotating HWP1, one can change the pump polarization as $\ket{H} \longrightarrow \cos\theta \ket{H} + \sin\theta \ket{V}$, where $H$ and $V$ stand for horizontal and vertical polarization.  By tilting QWP1 one can adjust the relative phase \cite{kwiat99}.   The pump polarization is then $\ket{H} \longrightarrow \cos\theta \ket{H} + e^{i\phi}\sin\theta \ket{V}$ up to a global phase.   The pump beam then enters a polarization-dependent Sagnac interferometer with a nested Dove prism (DP) orientated at $22.5^{\circ}$.  This type of Sagnac interferometer is experimentally advantageous since it is insensitive to phase fluctuations, and has been used to construct an optical single-photon CNOT gate \cite{fiorentino04} and to measure the spatial Wigner function \cite{mukamel03}.  It is well known that a Dove prism orientated at an azimuthal angle $\varphi$ rotates an image by an angle $2\varphi$ in the transverse plane.  The polarizing beam splitter sends $H$ and $V$-polarized components into opposite ends of the interferometer, where the Dove prism rotates the image of the $H$-polarized component by $45^{\circ}$, while the $V$-polarized component, which is propagating in the opposite direction, is rotated by $-45^{\circ}$.  The second Dove prism (located outside the interferometer) is used to realign the images in the horizontal-vertical coordinate system.  A Dove prism will also slightly rotate the polarization direction.  However, since in all cases the Dove prisms are followed by PBS's which project onto the desired polarization direction, this will result in only a slight reduction in beam intensity.  After the second Dove prism, the pump beam is  in a superposition:  $\cos\theta \ket{H}\mathcal{V}_{02} + e^{i\phi}\sin\theta \ket{V}\mathcal{V}_{20}$.  Using a polarizing beam splitter (PBS), one can project onto the $45^{\circ}$-polarization component, after which the pump beam is in the superposition: $\ket{45}(\cos\theta \mathcal{V}_{02} + e^{i\phi}\sin\theta \mathcal{V}_{20})$.  The last half-wave plate (HWP2) is used to realign the polarization before entering the non-linear crystal, where the HG components generate the terms in table \ref{tab:hg02}.  Post-selecting the first-order terms ($j+k=1,u+t=1$), the two photon state is
 \begin{equation}
   \ket{\psi} = \cos\theta\ket{v_{01}}\ket{v_{01}} + e^{i\phi}\sin\theta\ket{v_{10}}\ket{v_{10}}
   \label{eq:nonmaxstate}
   \end{equation}
  The weights and relative phase of the two-photon state (\ref{eq:nonmaxstate}) can be adjusted by rotating HWP1 and tilting QWP1, which, along with rotations and reflections of the down-converted fields described in the last section, allows for the creation of any bipartite pure state.
 \subsection{Hyperentangled states}
 Another interesting possibility is the creation of hyperentangled states \cite{kwiat97}.  It has been shown that such states may be useful in quantum dense coding and quantum cryptography \cite{walborn03c}.  Using one of the experimental situations described above and replacing the single type-I crystal we have considered with either the type-II ``crossed cone" source \cite{kwiat95} or the two-crystal type-I source \cite{kwiat99} of polarization-entangled photons, it should be possible to generate a two-photon state entangled in HG-mode and polarization.  These sources generally require that the crystals are thin (on the order of a few millimeters), so the possible astigmatism effects discussed  in section \ref{sec:gen} should be minimal, even for the type-II source.  Moreover, the experimental setups described above require only lower order HG modes.
 \section{Conclusion}
 We have shown that it is possible to generate correlated Hermite-Gaussian modes through spontaneous parametric down-conversion.  We have derived exact and approximate analytical expressions for the probability amplitudes $C^{\,nm}_{jkut}$ to generate arbitrary combinations of Hermite-Gaussian fields.  For any Hermite-Gaussian pump beam, there exist parity conservation conditions for the $x$ and $y$ indices of the down-converted Hermite-Gaussian modes.  We have used these results to show that the two-photon state is indeed entangled in Hermite-Gaussian modes.  We have discussed the generation of maximally entangled Bell-states and non-maximally entangled pure states of first-order Hermite-Gaussian fields.  These results can be used to engineer entangled states of higher-dimension, and promise to be useful in quantum information schemes.
 \appendix*
 \section{Calculation of $C^{\,nm}_{jkut}$}
 In section \ref{sec:gen}, we showed that the coefficient
 $C^{\,nm}_{jkut}$is given by Eq. (\ref{eq:hgspdc7}):
\begin{align}
C^{\,nm}_{jkut} = & \frac{1}{\pi}\sqrt{\frac{2 L}{K}}
\iint    d\rmvect{q}_{s}\,d\rmvect{q}_{i}
v^{*}_{jk}(\rmvect{q}_{s})v^{*}_{ut}(\rmvect{q}_{i}) \times
\nonumber \\ & \mathcal{V}_{nm}(\rmvect{q}_{s}+\rmvect{q}_{i})\,
\sinc \left(\frac{L}{4K}|\rmvect{q}_{s}-\rmvect{q}_{i}|^2 \right
).
\end{align}
Changing coordinates to
\begin{align}
\rmvect{Q} = &  \rmvect{q}_{s} + \rmvect{q}_{i}, \nonumber \\
\rmvect{P} = &  \rmvect{q}_{s} - \rmvect{q}_{i},
\end{align}
such that $d\rmvect{q}_{s}\,d\rmvect{q}_{i} = d\rmvect{Q}\,d\rmvect{P}/2$, we have
\begin{align}
C^{\,nm}_{jkut} =  \frac{1}{\pi}\sqrt{\frac{L}{2 K}} \iint & d\rmvect{Q} d\rmvect{P} v^{*}_{jk}\left(\frac{\rmvect{Q}+\rmvect{P}}{2}\right)v^{*}_{ut}\left(\frac{\rmvect{Q}-\rmvect{P}}{2}\right) \times \nonumber \\
& \mathcal{V}_{nm}(\rmvect{Q})\,
\sinc \left(\frac{L}{4K}P^2 \right ).
\label{eq:hgspdc8}
\end{align}
\par
Now consider a down-converted HG mode $v_{nm}$ with wavelength $\lambda_{c}$ and beam radius $w_{0c}$.  To be more precise, let us temporarily write $v_{nm}(\rmvect{q},\lambda_{c},w_{0c})$.  Since we are working with down-converted fields satisfying $\lambda_{c}=2\lambda_{p}$, it is easy to show from the general form of HG modes that $v_{nm}(\rmvect{q}/\sqrt{2},\lambda_{c},\sqrt{2}w_{0p}) = \mathcal{V}_{nm}(\rmvect{q},\lambda_{p},w_{0p})$.  That is, the down-converted HG modes with $w_{0c}=\sqrt{2}w_{0p}$ will have the same Rayleigh range $z_{R}$, Gouy phase $\gouy(z)$ and radius of curvature $R(z)$ as the pump field.  Using this property of the Gaussian modes,
we can expand $v^{*}_{jk}((\rmvect{Q}+\rmvect{P})/{2})$ and $v^{*}_{ut}((\rmvect{Q}-\rmvect{P})/{2})$  and regroup the $x$ and $y$ terms, which gives
\begin{widetext}
\begin{equation}
v^{*}_{jk}\left(\frac{\rmvect{Q}+\rmvect{P}}{2}\right)v^{*}_{ut}\left(\frac{\rmvect{Q}-\rmvect{P}}{2}\right) =
  \mathcal{V}^{*}_{ju}\left(\frac{Q_{x}+P_{x}}{\sqrt{2}},\frac{Q_{x}-P_{x}}{\sqrt{2}} \right)
 \mathcal{V}^{*}_{kt}\left(\frac{Q_{y}+P_{y}}{\sqrt{2}},\frac{Q_{y}-P_{y}}{\sqrt{2}} \right),
\label{eq:hgspdc12}
\end{equation}
where we used definitions (\ref{eq:Cnm}) and (\ref{eq:Dnm}) to show that $D^{*}_{jk}D^{*}_{ut} =  D^{*}_{ju}D^{*}_{kt}$.  We note here that relation \eqref{eq:hgspdc12} is valid for all $z_{s}=z_{i}$.  Then, using the definition of the DHG modes (\ref{eq:DHGexpan}), Eq. (\ref{eq:hgspdc12}) can be expressed as
\begin{equation}
v^{*}_{jk}\left(\frac{\rmvect{Q}+\rmvect{P}}{2}\right)v^{*}_{ut}\left(\frac{\rmvect{Q}-\rmvect{P}}{2}\right) =
  \sum_{\alpha=0}^{N}b(j,u,\alpha)\mathcal{V}^{*}_{N-\alpha,\alpha}\left(Q_{x},P_{x} \right)
  \sum_{\beta=0}^{M}b(k,t,\beta)\mathcal{V}^{*}_{M-\beta,\beta}\left(Q_{y},P_{y} \right),
\label{eq:hgspdc13}
\end{equation}
where $N=j+u$ and $M=k+t$.
Noting that $D^{*}_{N-\alpha,\alpha}D^{*}_{M-\beta,\beta} = D^{*}_{N-\alpha,M-\beta}D^{*}_{\alpha,\beta}$, it is straightforward to show that the product of angular spectra on the RHS of Eq. (\ref{eq:hgspdc13}) can be rewritten as
\begin{equation}
\mathcal{V}^{*}_{N-\alpha,\alpha}\left(Q_{x},P_{x} \right)
\mathcal{V}^{*}_{M-\beta,\beta}\left(Q_{y},P_{y} \right)
=  \mathcal{V}^{*}_{N-\alpha,M-\beta}(\rmvect{Q})\mathcal{V}^{*}_{\alpha,\beta}(\rmvect{P}).
\label{eq:14}
\end{equation}
  Putting these Eqs. (\ref{eq:hgspdc13}) and (\ref{eq:14}) back into Eq. \eqref{eq:hgspdc8}, the coefficent $C^{\,nm}_{jkut}$ becomes
\begin{equation}
C^{\,nm}_{jkut} = \frac{1}{\pi}\sqrt{\frac{L}{2 K}} \sum_{\alpha=0}^{N}\sum_{\beta=0}^{M}b(j,u,\alpha)b(k,t,\beta)   \int d\rmvect{Q}\, \mathcal{V}^{*}_{N-\alpha,M-\beta}(\rmvect{Q}) \mathcal{V}_{nm}(\rmvect{Q})  \int d\rmvect{P}\,\mathcal{V}^{*}_{\alpha,\beta}(\rmvect{P})\,
\sinc \left(\frac{L}{4K}P^2 \right ).
\label{eq:hgspdc15}
\end{equation}
\end{widetext}
\par
The HG modes are orthonormal, so
\begin{equation}
\int d\rmvect{Q}\, \mathcal{V}^{*}_{N-\alpha,M-\beta}(\rmvect{Q}) \mathcal{V}_{nm}(\rmvect{Q}) = \delta_{N-\alpha,n}\delta_{M-\beta,m}.
\end{equation}
which gives
\begin{align}
C^{\,nm}_{jkut} = & \frac{1}{\pi}\sqrt{\frac{L}{2 K}}  b(j,u,N-n)b(k,t,M-m)  \nonumber \\
& \int d\rmvect{P}\,\mathcal{V}^{*}_{N-n,M-m}(\rmvect{P})\,
\sinc \left(\frac{L}{4K}P^2 \right ),
\label{eq:hgspdc16}
\end{align}
if $N=j+u \geq n$ and $M=k+t \geq m$, else $C^{\,nm}_{jkut} = 0$.
\par
Using the following expression for the Hermite polynomials \cite{lebedev72}:
\begin{equation}
H_{n}(\xi) = \sum\limits_{j=0}^{n/2}\frac{(-1)^{j}n!}{j!(n-2j)!}(2\xi)^{n-2j},
\end{equation}
it is straightforward to calculate the integral in (\ref{eq:hgspdc16}) analytically.   After some algebraic manipulation,
\begin{align}
C^{\,nm}_{jkut} = & \sqrt{\frac{\alpha!\beta! }{A \pi}}\left( \frac{1}{2} \right)^{\frac{\alpha+\beta}{2}} \frac{\arctan A}{(\alpha/2)!(\beta/2)!} b(j,u,\alpha)b(k,t,\beta) \times \nonumber \\
& \sum\limits_{r=0}^{(\alpha+\beta)/2}
\left (\begin{array}{c}
 \frac{\alpha+\beta}{2} \\  r
\end{array} \right)
\left( \frac{-2}{\sqrt{1+A^2}}\right)^{r} \sinc ( r \arctan A)
\end{align}
if $j+u \geq n$ and $k+t \geq m$, else $C^{\,nm}_{jkut} = 0$.  Here we have defined $A=L/Kw_{p}^2$, $\alpha=N-n$, $\beta=M-m$ and used the usual binomial coefficient.
 \par
For thin non-linear crystals, it is possible to arrive at a more revealing solution.  If the nonlinear crystal is thin ($L \sim 1$ mm), we can approximate $\sinc({L}/{4K}P^2 ) \approx 1$ in equation (\ref{eq:hgspdc16}), giving $\int d \rmvect{P} \mathcal{V}_{\alpha,\beta}^{*}(\rmvect{P}) \sinc({L}/{4K}P^2 ) \approx \int d \rmvect{P} \mathcal{V}_{\alpha,\beta}^{*}(\rmvect{P})$.  Numerical integration shows that errors due to this approximation are less than 3\% for modes as high as $\alpha=\beta=10$ for typical experimental values.   Then, integration in $\rmvect{P}$ transforms (\ref{eq:Cexact}) to
 \begin{align}
{C}^{\,nm}_{jkut}\rightarrow {\tilde{C}}^{\,nm}_{jkut} =
  & \sqrt{\frac{2 A}{\pi}}b(j,u,N-n)b(k,t,M-m) \times \nonumber \\
  & \HG_{N-n,M-m}(0,0,0),
   \end{align}
 if $j+u \geq n$ and $k+t \geq m$, otherwise $C^{\,nm}_{jkut} = 0$.    $\HG_{\gamma\delta}$ is the Hermite-Gaussian mode (\ref{eq:hg}) evaluated at $x=y=z=0$.
  \begin{acknowledgments}
The authors acknowledge financial support from the Brazilian funding agencies CNPq, CAPES and the Milenium Institute for Quantum
Information.  We thank A. G. Costa Moura for useful discussions.
\end{acknowledgments}

\begin{thebibliography}{35}
\expandafter\ifx\csname natexlab\endcsname\relax\def\natexlab#1{#1}\fi
\expandafter\ifx\csname bibnamefont\endcsname\relax
  \def\bibnamefont#1{#1}\fi
\expandafter\ifx\csname bibfnamefont\endcsname\relax
  \def\bibfnamefont#1{#1}\fi
\expandafter\ifx\csname citenamefont\endcsname\relax
  \def\citenamefont#1{#1}\fi
\expandafter\ifx\csname url\endcsname\relax
  \def\url#1{\texttt{#1}}\fi
\expandafter\ifx\csname urlprefix\endcsname\relax\def\urlprefix{URL }\fi
\providecommand{\bibinfo}[2]{#2}
\providecommand{\eprint}[2][]{\url{#2}}

\bibitem[{\citenamefont{Saleh and Teich}(1991)}]{saleh91}
\bibinfo{author}{\bibfnamefont{B.~E.~A.} \bibnamefont{Saleh}} \bibnamefont{and}
  \bibinfo{author}{\bibfnamefont{M.~C.} \bibnamefont{Teich}},
  \emph{\bibinfo{title}{Fundamental Photonics}} (\bibinfo{publisher}{Wiley},
  \bibinfo{address}{New York}, \bibinfo{year}{1991}).

\bibitem[{\citenamefont{Allen et~al.}(1992)\citenamefont{Allen, Beijersbergen,
  Spreeuw, and Woerdman}}]{allen92}
\bibinfo{author}{\bibfnamefont{L.}~\bibnamefont{Allen}},
  \bibinfo{author}{\bibfnamefont{M.~W.} \bibnamefont{Beijersbergen}},
  \bibinfo{author}{\bibfnamefont{R.~J.~C.} \bibnamefont{Spreeuw}},
  \bibnamefont{and} \bibinfo{author}{\bibfnamefont{J.~P.}
  \bibnamefont{Woerdman}}, \bibinfo{journal}{Phys. Rev. A}
  \textbf{\bibinfo{volume}{45}}, \bibinfo{pages}{8185} (\bibinfo{year}{1992}).

\bibitem[{\citenamefont{van Enk and Nienhuis}(1992)}]{vanenk92}
\bibinfo{author}{\bibfnamefont{S.~J.} \bibnamefont{van Enk}} \bibnamefont{and}
  \bibinfo{author}{\bibfnamefont{G.}~\bibnamefont{Nienhuis}},
  \bibinfo{journal}{Optics Comm.} \textbf{\bibinfo{volume}{94}},
  \bibinfo{pages}{147} (\bibinfo{year}{1992}).

\bibitem[{\citenamefont{Heckenberg et~al.}(1992)\citenamefont{Heckenberg,
  McDuff, Smith, and White}}]{heckenberg92}
\bibinfo{author}{\bibfnamefont{N.~R.} \bibnamefont{Heckenberg}},
  \bibinfo{author}{\bibfnamefont{R.}~\bibnamefont{McDuff}},
  \bibinfo{author}{\bibfnamefont{C.~P.} \bibnamefont{Smith}}, \bibnamefont{and}
  \bibinfo{author}{\bibfnamefont{A.~G.} \bibnamefont{White}},
  \bibinfo{journal}{Optics Letters} \textbf{\bibinfo{volume}{17}},
  \bibinfo{pages}{221} (\bibinfo{year}{1992}).

\bibitem[{\citenamefont{Leach et~al.}(2002)\citenamefont{Leach, Padgett,
  Barnett, Franke-Arnold, and Courtial}}]{leach02}
\bibinfo{author}{\bibfnamefont{J.}~\bibnamefont{Leach}},
  \bibinfo{author}{\bibfnamefont{M.~J.} \bibnamefont{Padgett}},
  \bibinfo{author}{\bibfnamefont{S.~M.} \bibnamefont{Barnett}},
  \bibinfo{author}{\bibfnamefont{S.}~\bibnamefont{Franke-Arnold}},
  \bibnamefont{and} \bibinfo{author}{\bibfnamefont{J.}~\bibnamefont{Courtial}},
  \bibinfo{journal}{Phys. Rev. Lett.} \textbf{\bibinfo{volume}{88}},
  \bibinfo{pages}{257901} (\bibinfo{year}{2002}).

\bibitem[{\citenamefont{Wei et~al.}(2003)\citenamefont{Wei, Xue, Leach,
  Padgett, Barnett, Franke-Arnold, Yao, and Courtial}}]{wei03_OC117}
\bibinfo{author}{\bibfnamefont{H.}~\bibnamefont{Wei}},
  \bibinfo{author}{\bibfnamefont{X.}~\bibnamefont{Xue}},
  \bibinfo{author}{\bibfnamefont{J.}~\bibnamefont{Leach}},
  \bibinfo{author}{\bibfnamefont{M.~J.} \bibnamefont{Padgett}},
  \bibinfo{author}{\bibfnamefont{S.~M.} \bibnamefont{Barnett}},
  \bibinfo{author}{\bibfnamefont{S.}~\bibnamefont{Franke-Arnold}},
  \bibinfo{author}{\bibfnamefont{E.}~\bibnamefont{Yao}}, \bibnamefont{and}
  \bibinfo{author}{\bibfnamefont{J.}~\bibnamefont{Courtial}},
  \bibinfo{journal}{Optics Comm.} \textbf{\bibinfo{volume}{223}},
  \bibinfo{pages}{117} (\bibinfo{year}{2003}).

\bibitem[{\citenamefont{Mair et~al.}(2001)\citenamefont{Mair, Vaziri, Weihs,
  and Zeilinger}}]{mair01}
\bibinfo{author}{\bibfnamefont{A.}~\bibnamefont{Mair}},
  \bibinfo{author}{\bibfnamefont{A.}~\bibnamefont{Vaziri}},
  \bibinfo{author}{\bibfnamefont{G.}~\bibnamefont{Weihs}}, \bibnamefont{and}
  \bibinfo{author}{\bibfnamefont{A.}~\bibnamefont{Zeilinger}},
  \bibinfo{journal}{Nature} \textbf{\bibinfo{volume}{412}},
  \bibinfo{pages}{313} (\bibinfo{year}{2001}).

\bibitem[{\citenamefont{Arnaut and Barbosa}(2001)}]{arnaut01}
\bibinfo{author}{\bibfnamefont{H.~H.} \bibnamefont{Arnaut}} \bibnamefont{and}
  \bibinfo{author}{\bibfnamefont{G.~A.} \bibnamefont{Barbosa}},
  \bibinfo{journal}{Phys. Rev. Lett.} \textbf{\bibinfo{volume}{85}},
  \bibinfo{pages}{286} (\bibinfo{year}{2001}).

\bibitem[{\citenamefont{Franke-Arnold et~al.}(2002)\citenamefont{Franke-Arnold,
  Barnett, Padgett, and Allen}}]{franke-arnold02}
\bibinfo{author}{\bibfnamefont{S.}~\bibnamefont{Franke-Arnold}},
  \bibinfo{author}{\bibfnamefont{S.~M.} \bibnamefont{Barnett}},
  \bibinfo{author}{\bibfnamefont{M.~J.} \bibnamefont{Padgett}},
  \bibnamefont{and} \bibinfo{author}{\bibfnamefont{L.}~\bibnamefont{Allen}},
  \bibinfo{journal}{Phys. Rev. A} \textbf{\bibinfo{volume}{65}},
  \bibinfo{pages}{033823} (\bibinfo{year}{2002}).

\bibitem[{\citenamefont{Barbosa and Arnaut}(2002)}]{barbosa02}
\bibinfo{author}{\bibfnamefont{G.~A.} \bibnamefont{Barbosa}} \bibnamefont{and}
  \bibinfo{author}{\bibfnamefont{H.~H.} \bibnamefont{Arnaut}},
  \bibinfo{journal}{Phys. Rev. A} \textbf{\bibinfo{volume}{65}},
  \bibinfo{pages}{053801} (\bibinfo{year}{2002}).

\bibitem[{\citenamefont{Torres et~al.}(2003)\citenamefont{Torres, Deyanova,
  Torner, and Molina-Terriza}}]{torres03_PRA052313}
\bibinfo{author}{\bibfnamefont{J.~P.} \bibnamefont{Torres}},
  \bibinfo{author}{\bibfnamefont{Y.}~\bibnamefont{Deyanova}},
  \bibinfo{author}{\bibfnamefont{L.}~\bibnamefont{Torner}}, \bibnamefont{and}
  \bibinfo{author}{\bibfnamefont{G.}~\bibnamefont{Molina-Terriza}},
  \bibinfo{journal}{Phys. Rev. A.} \textbf{\bibinfo{volume}{67}},
  \bibinfo{pages}{052313} (\bibinfo{year}{2003}).

\bibitem[{\citenamefont{Walborn et~al.}(2004)\citenamefont{Walborn,
  de~Oliveira, Thebaldi, and Monken}}]{walborn04a}
\bibinfo{author}{\bibfnamefont{S.~P.} \bibnamefont{Walborn}},
  \bibinfo{author}{\bibfnamefont{A.~N.} \bibnamefont{de~Oliveira}},
  \bibinfo{author}{\bibfnamefont{R.~S.} \bibnamefont{Thebaldi}},
  \bibnamefont{and} \bibinfo{author}{\bibfnamefont{C.~H.}
  \bibnamefont{Monken}}, \bibinfo{journal}{Phys. Rev. A}
  \textbf{\bibinfo{volume}{69}}, \bibinfo{pages}{023811}
  (\bibinfo{year}{2004}).

\bibitem[{\citenamefont{Ren et~al.}(2004)\citenamefont{Ren, Guo, Yu, Li, and
  Guo}}]{ren04}
\bibinfo{author}{\bibfnamefont{X.-F.} \bibnamefont{Ren}},
  \bibinfo{author}{\bibfnamefont{G.-P.} \bibnamefont{Guo}},
  \bibinfo{author}{\bibfnamefont{B.}~\bibnamefont{Yu}},
  \bibinfo{author}{\bibfnamefont{J.}~\bibnamefont{Li}}, \bibnamefont{and}
  \bibinfo{author}{\bibfnamefont{G.-C.} \bibnamefont{Guo}},
  \bibinfo{journal}{J. Opt. B: Quantum Semiclass. Opt.}
  \textbf{\bibinfo{volume}{6}}, \bibinfo{pages}{243} (\bibinfo{year}{2004}).

\bibitem[{\citenamefont{Law and Eberly}(2004)}]{law04}
\bibinfo{author}{\bibfnamefont{C.~K.} \bibnamefont{Law}} \bibnamefont{and}
  \bibinfo{author}{\bibfnamefont{J.~H.} \bibnamefont{Eberly}},
  \bibinfo{journal}{Phys. Rev. Lett.} \textbf{\bibinfo{volume}{92}},
  \bibinfo{pages}{127903} (\bibinfo{year}{2004}).

\bibitem[{\citenamefont{Vaziri et~al.}(2002)\citenamefont{Vaziri, Weihs, and
  Zeilinger}}]{vaziri02}
\bibinfo{author}{\bibfnamefont{A.}~\bibnamefont{Vaziri}},
  \bibinfo{author}{\bibfnamefont{G.}~\bibnamefont{Weihs}}, \bibnamefont{and}
  \bibinfo{author}{\bibfnamefont{A.}~\bibnamefont{Zeilinger}},
  \bibinfo{journal}{Phys. Rev. Lett.} \textbf{\bibinfo{volume}{89}},
  \bibinfo{pages}{240401} (\bibinfo{year}{2002}).

\bibitem[{\citenamefont{Langford et~al.}(2004)\citenamefont{Langford, Dalton,
  Harvey, O'Brien, Pryde, Gilchrist, Bartlett, and White}}]{langford03}
\bibinfo{author}{\bibfnamefont{N.~K.} \bibnamefont{Langford}},
  \bibinfo{author}{\bibfnamefont{R.~B.} \bibnamefont{Dalton}},
  \bibinfo{author}{\bibfnamefont{M.~D.} \bibnamefont{Harvey}},
  \bibinfo{author}{\bibfnamefont{J.~L.} \bibnamefont{O'Brien}},
  \bibinfo{author}{\bibfnamefont{G.~J.} \bibnamefont{Pryde}},
  \bibinfo{author}{\bibfnamefont{A.}~\bibnamefont{Gilchrist}},
  \bibinfo{author}{\bibfnamefont{S.~D.} \bibnamefont{Bartlett}},
  \bibnamefont{and} \bibinfo{author}{\bibfnamefont{A.~G.} \bibnamefont{White}},
  \bibinfo{journal}{Phys. Rev. Lett.} \textbf{\bibinfo{volume}{93}},
  \bibinfo{pages}{053601} (\bibinfo{year}{2004}).

\bibitem[{\citenamefont{O'Neil and Courtial}(2000)}]{oneil00}
\bibinfo{author}{\bibfnamefont{A.~T.} \bibnamefont{O'Neil}} \bibnamefont{and}
  \bibinfo{author}{\bibfnamefont{J.}~\bibnamefont{Courtial}},
  \bibinfo{journal}{Opt. Comm.} \textbf{\bibinfo{volume}{181}},
  \bibinfo{pages}{35} (\bibinfo{year}{2000}).

\bibitem[{\citenamefont{Xue et~al.}(2001)\citenamefont{Xue, Wei, and
  Kirk}}]{xue01_OL}
\bibinfo{author}{\bibfnamefont{X.}~\bibnamefont{Xue}},
  \bibinfo{author}{\bibfnamefont{H.}~\bibnamefont{Wei}}, \bibnamefont{and}
  \bibinfo{author}{\bibfnamefont{A.~G.} \bibnamefont{Kirk}},
  \bibinfo{journal}{Optics Letters} \textbf{\bibinfo{volume}{26}},
  \bibinfo{pages}{1746} (\bibinfo{year}{2001}).

\bibitem[{\citenamefont{Sasada and Okamoto}(2003)}]{sasada03_PRA012323}
\bibinfo{author}{\bibfnamefont{H.}~\bibnamefont{Sasada}} \bibnamefont{and}
  \bibinfo{author}{\bibfnamefont{M.}~\bibnamefont{Okamoto}},
  \bibinfo{journal}{Phys. Rev. A.} \textbf{\bibinfo{volume}{68}},
  \bibinfo{pages}{012323} (\bibinfo{year}{2003}).

\bibitem[{\citenamefont{Beijersbergen et~al.}(1993)\citenamefont{Beijersbergen,
  Allen, van~der Veen, and Woerdman}}]{beijersbergen93}
\bibinfo{author}{\bibfnamefont{M.~W.} \bibnamefont{Beijersbergen}},
  \bibinfo{author}{\bibfnamefont{L.}~\bibnamefont{Allen}},
  \bibinfo{author}{\bibfnamefont{H.~E. L.~O.} \bibnamefont{van~der Veen}},
  \bibnamefont{and} \bibinfo{author}{\bibfnamefont{J.~P.}
  \bibnamefont{Woerdman}}, \bibinfo{journal}{Optics Comm.}
  \textbf{\bibinfo{volume}{96}}, \bibinfo{pages}{123} (\bibinfo{year}{1993}).

\bibitem[{\citenamefont{Hong and Mandel}(1985)}]{hong85}
\bibinfo{author}{\bibfnamefont{C.~K.} \bibnamefont{Hong}} \bibnamefont{and}
  \bibinfo{author}{\bibfnamefont{L.}~\bibnamefont{Mandel}},
  \bibinfo{journal}{Phys. Rev. A} \textbf{\bibinfo{volume}{31}},
  \bibinfo{pages}{2409} (\bibinfo{year}{1985}).

\bibitem[{\citenamefont{Monken et~al.}(1998)\citenamefont{Monken, Souto Ribeiro, and
  P\'adua}}]{monken98a}
\bibinfo{author}{\bibfnamefont{C.~H.} \bibnamefont{Monken}},
  \bibinfo{author}{\bibfnamefont{P.~H.} \bibnamefont{Souto Ribeiro}},
  \bibnamefont{and} \bibinfo{author}{\bibfnamefont{S.}~\bibnamefont{P\'adua}},
  \bibinfo{journal}{Phys. Rev. A.} \textbf{\bibinfo{volume}{57}},
  \bibinfo{pages}{3123} (\bibinfo{year}{1998}).

\bibitem[{\citenamefont{Kwiat et~al.}(1995)\citenamefont{Kwiat, Mattle,
  Weinfurter, Zeilinger, Sergienko, and Shih}}]{kwiat95}
\bibinfo{author}{\bibfnamefont{P.~G.} \bibnamefont{Kwiat}},
  \bibinfo{author}{\bibfnamefont{K.}~\bibnamefont{Mattle}},
  \bibinfo{author}{\bibfnamefont{H.}~\bibnamefont{Weinfurter}},
  \bibinfo{author}{\bibfnamefont{A.}~\bibnamefont{Zeilinger}},
  \bibinfo{author}{\bibfnamefont{A.~V.} \bibnamefont{Sergienko}},
  \bibnamefont{and} \bibinfo{author}{\bibfnamefont{Y.}~\bibnamefont{Shih}},
  \bibinfo{journal}{Phys. Rev. Lett.} \textbf{\bibinfo{volume}{75}},
  \bibinfo{pages}{4337} (\bibinfo{year}{1995}).

\bibitem[{\citenamefont{Atat\"ure et~al.}(2002)\citenamefont{Atat\"ure,
  Di Giuseppe, Shaw, Sergienko, Saleh, and Teich}}]{atature02_PRA23822}
\bibinfo{author}{\bibfnamefont{M.}~\bibnamefont{Atat\"ure}},
  \bibinfo{author}{\bibfnamefont{G.} \bibnamefont{Di Giuseppe}},
  \bibinfo{author}{\bibfnamefont{M.~D.} \bibnamefont{Shaw}},
  \bibinfo{author}{\bibfnamefont{A.~V.} \bibnamefont{Sergienko}},
  \bibinfo{author}{\bibfnamefont{B.~E.~A.} \bibnamefont{Saleh}},
  \bibnamefont{and} \bibinfo{author}{\bibfnamefont{M.~C.} \bibnamefont{Teich}},
  \bibinfo{journal}{Phys. Rev. A} \textbf{\bibinfo{volume}{66}},
  \bibinfo{pages}{023822} (\bibinfo{year}{2002}).

\bibitem[{\citenamefont{Moura and Monken}()}]{costamoura04}
\bibinfo{author}{\bibfnamefont{A.~G.~C.} \bibnamefont{Moura}} \bibnamefont{and}
  \bibinfo{author}{\bibfnamefont{C.~H.} \bibnamefont{Monken}},
  \bibinfo{note}{in preparation}.

\bibitem[{\citenamefont{Nielsen and Chuang}(2000)}]{chuang00}
\bibinfo{author}{\bibfnamefont{M.}~\bibnamefont{Nielsen}} \bibnamefont{and}
  \bibinfo{author}{\bibfnamefont{I.}~\bibnamefont{Chuang}},
  \emph{\bibinfo{title}{Quantum Computation and Quantum Information}}
  (\bibinfo{publisher}{Cambridge}, \bibinfo{address}{Cambridge},
  \bibinfo{year}{2000}).

\bibitem[{\citenamefont{Ballentine}(1998)}]{ballentine98}
\bibinfo{author}{\bibfnamefont{L.}~\bibnamefont{Ballentine}},
  \emph{\bibinfo{title}{Quantum Mechanics: A Modern Developement}}
  (\bibinfo{publisher}{World Scientific}, \bibinfo{address}{Singapore},
  \bibinfo{year}{1998}).

\bibitem[{\citenamefont{Kantorovich and Akilov}(1982)}]{kantorovich82}
\bibinfo{author}{\bibfnamefont{L.~V.} \bibnamefont{Kantorovich}}
  \bibnamefont{and} \bibinfo{author}{\bibfnamefont{G.~P.}
  \bibnamefont{Akilov}}, \emph{\bibinfo{title}{Functional Analysis}}
  (\bibinfo{publisher}{Pergamon Press}, \bibinfo{address}{England},
  \bibinfo{year}{1982}).

\bibitem[{\citenamefont{Vaziri et~al.}(2003)\citenamefont{Vaziri, Pan,
  Jennewein, Weihs, and Zeilinger}}]{vaziri03a}
\bibinfo{author}{\bibfnamefont{A.}~\bibnamefont{Vaziri}},
  \bibinfo{author}{\bibfnamefont{J.-W.} \bibnamefont{Pan}},
  \bibinfo{author}{\bibfnamefont{T.}~\bibnamefont{Jennewein}},
  \bibinfo{author}{\bibfnamefont{G.}~\bibnamefont{Weihs}}, \bibnamefont{and}
  \bibinfo{author}{\bibfnamefont{A.}~\bibnamefont{Zeilinger}},
  \bibinfo{journal}{Phys. Rev. Lett.} \textbf{\bibinfo{volume}{91}},
  \bibinfo{pages}{227902} (\bibinfo{year}{2003}).

\bibitem[{\citenamefont{Kwiat et~al.}(1999)\citenamefont{Kwiat, Waks, White,
  Appelbaum, and Eberhard}}]{kwiat99}
\bibinfo{author}{\bibfnamefont{P.~G.} \bibnamefont{Kwiat}},
  \bibinfo{author}{\bibfnamefont{E.}~\bibnamefont{Waks}},
  \bibinfo{author}{\bibfnamefont{A.~G.} \bibnamefont{White}},
  \bibinfo{author}{\bibfnamefont{I.}~\bibnamefont{Appelbaum}},
  \bibnamefont{and} \bibinfo{author}{\bibfnamefont{P.~H.}
  \bibnamefont{Eberhard}}, \bibinfo{journal}{Phys. Rev. A.}
  \textbf{\bibinfo{volume}{60}}, \bibinfo{pages}{R773} (\bibinfo{year}{1999}).

\bibitem[{\citenamefont{Fiorentino and Wong}(2004)}]{fiorentino04}
\bibinfo{author}{\bibfnamefont{M.}~\bibnamefont{Fiorentino}} \bibnamefont{and}
  \bibinfo{author}{\bibfnamefont{F.~N.~C.} \bibnamefont{Wong}},
  \bibinfo{journal}{Phys. Rev. Lett.} \textbf{\bibinfo{volume}{93}},
  \bibinfo{pages}{070502} (\bibinfo{year}{2004}).

\bibitem[{\citenamefont{Mukamel et~al.}(2003)\citenamefont{Mukamel, Banaszek,
  Walmsley, and Dorrer}}]{mukamel03}
\bibinfo{author}{\bibfnamefont{E.}~\bibnamefont{Mukamel}},
  \bibinfo{author}{\bibfnamefont{K.}~\bibnamefont{Banaszek}},
  \bibinfo{author}{\bibfnamefont{I.~A.} \bibnamefont{Walmsley}},
  \bibnamefont{and} \bibinfo{author}{\bibfnamefont{C.}~\bibnamefont{Dorrer}},
  \bibinfo{journal}{Opt. Lett.} \textbf{\bibinfo{volume}{28}},
  \bibinfo{pages}{1317} (\bibinfo{year}{2003}).

\bibitem[{\citenamefont{Kwiat}(1997)}]{kwiat97}
\bibinfo{author}{\bibfnamefont{P.~G.} \bibnamefont{Kwiat}},
  \bibinfo{journal}{J. Mod. Optics} \textbf{\bibinfo{volume}{44}},
  \bibinfo{pages}{2173} (\bibinfo{year}{1997}).

\bibitem[{\citenamefont{Walborn et~al.}(2003)\citenamefont{Walborn, P\'adua,
  and Monken}}]{walborn03c}
\bibinfo{author}{\bibfnamefont{S.~P.} \bibnamefont{Walborn}},
  \bibinfo{author}{\bibfnamefont{S.}~\bibnamefont{P\'adua}}, \bibnamefont{and}
  \bibinfo{author}{\bibfnamefont{C.~H.} \bibnamefont{Monken}},
  \bibinfo{journal}{Phys. Rev. A} \textbf{\bibinfo{volume}{68}},
  \bibinfo{pages}{042313} (\bibinfo{year}{2003}).

\bibitem[{\citenamefont{Lebedev}(1972)}]{lebedev72}
\bibinfo{author}{\bibfnamefont{N.~N.} \bibnamefont{Lebedev}},
  \emph{\bibinfo{title}{Special Functions and Their Applications}}
  (\bibinfo{publisher}{Dover}, \bibinfo{address}{New York},
  \bibinfo{year}{1972}).

\end{thebibliography}

\end{document}